# BPI: A Novel Efficient and Reliable Search Structure for Hybrid Storage Blockchain

Xinkui Zhao
School of Software Technology,
Zhejiang University
Ningbo, China
zhaoxinkui@zju.edu.cn

Rengrong Xiong
School of Software Technology,
Zhejiang University
Ningbo, China
863899787@qq.com

Guanjie Cheng*
School of Software Technology,
Zhejiang University
Ningbo, China
chengguanjie@zju.edu.cn

Xinhao Jin
School of Software Technology,
Zhejiang University
Ningbo, China
2040006204@qq.com

Shawn Shi
School of Software Technology,
Zhejiang University
Ningbo, China
3396746834@qq.com

Xiubo Liang*
School of Software Technology,
Zhejiang University
Ningbo, China
xiubo@zju.edu.cn

Gongsheng Yuan
School of Software Technology,
Zhejiang University
Ningbo, China
ygs@zju.edu.cn

Xiaoye Miao
School of Software Technology,
Zhejiang University
Ningbo, China
miaoxy@zju.edu.cn

Jianwei Yin
School of Computer Science and
Technology, Zhejiang University
Hangzhou, China
zjuyjw@cs.zju.edu.cn

Shuiguang Deng*
School of Computer Science and
Technology, Zhejiang University
Hangzhou, China
dengsg@zju.edu.cn

## ABSTRACT

Hybrid storage solutions have emerged as potent strategies to alleviate the data storage bottlenecks prevalent in blockchain systems. These solutions harness off-chain Storage Services Providers (SPs) in conjunction with Authenticated Data Structures (ADS) to ensure data integrity and accuracy. Despite these advancements, the reliance on centralized SPs raises concerns about query correctness. Although ADS can verify the existence of individual query results, they fall short of preventing SPs from omitting valid results.

In this paper, we delineate the fundamental distinctions between data search in blockchains and traditional database systems. Drawing upon these insights, we introduce BPI, a lightweight framework that enables efficient keyword queries and maintenance with low overhead. We propose "Articulated Search", a query pattern specifically designed for blockchain environments that enhances search efficiency while significantly reducing costs during data user updates. Furthermore, BPI employs a suite of validation models to ensure the inclusion of all valid content in search results while maintaining low overhead.

Extensive experimental evaluations demonstrate that the BPI framework achieves outstanding scalability and performance in keyword searches within blockchain, surpassing EthMB+ and state of the art search databases commonly used in mainstream hybrid storage blockchains (HSB).

**Note: This work has been accepted to ACM SIGMOD 2026.**

## 1 INTRODUCTION

With the rapid adoption of blockchain in academia and industry, it is widely used for secure, immutable, and transparent distributed data storage [15]. However, scalability remains a challenge due to constraints such as block size limits (e.g., Bitcoin's 1 MB cap) [39], latency issues from transaction validation and state synchronization [50]. Hybrid storage blockchains (HSBs) address this by integrating on-chain and off-chain storage [1, 8, 23], outsourcing data to SPs while retaining metadata on-chain to enhance scalability and reduce costs [49, 54].

Data search is a fundamental operation performed on HSB. Advanced search structures have been developed to support functionalities such as range searches [20, 45], keyword-based searches [26], and query result verification [3, 10, 21, 44]. For query result verification, existing approaches often rely on ADS maintained by the blockchain, with Merkle proofs commonly used to provide membership verification. However, in the HSB system, a malicious SP can undermine the completeness of query results while still passing blockchain verification. For example, an SP might return only a subset of the correct query results or provide data that exists in the database but does not satisfy the query conditions. To mitigate these vulnerabilities, the blockchain must execute query requests itself to ensure trusted and accurate verification results.

Consequently, developing cost-effective and efficient methods for on-chain query result verification becomes a critical challenge.

Existing search structures in HSBs, typically leveraging B+ trees [10], perform well in static environments but face challenges in dynamic blockchain contexts, characterized by frequent transactions that increase maintenance costs. For instance, Ethereum processes over 1.15 million transactions daily [11], resulting in high maintenance overhead for B+ tree-based index structures. Such overheads may be negligible for SPs with inexpensive resources, it poses significant challenges for blockchain nodes with costlier resources. Consequently, a novel data search structure capable of efficiently managing frequent updates while minimizing maintenance costs is required.

Furthermore, traditional databases typically support Create, Read, Update, and Delete (CRUD) operations, whereas blockchain ledger data is immutable, permitting only Create and Read (CR) operations (the discussion of updates here is limited to blockchain ledger data and differs from the later discussion of index structure updates). This immutability creates different search requirements, allowing users to skip previously queried data and focus on new data. In contrast, B+ tree queries introduce unnecessary overhead because they scan the entire tree. Even with secondary indexes, table lookup operations still incur redundant overhead, while creating additional indexes proportionally increases storage and insertion costs. Many existing studies on append-only databases effectively address this problem, such as retrieval structures like EASL [30], which leverages skip lists. EASL demonstrates substantial advantages when a time-increasing dimension serves as the primary key. However, it is ineffective for supporting keywords that do not follow temporal order. Constructing such indexes still requires "table lookup" and results in additional insertion overhead. The immutability feature is particularly crucial for applications such as supply chain management, where maintaining a consistent and tamper-evident record of transactions is essential to track goods [35]. Despite its significance, this distinctive characteristic of blockchain data has not been adequately explored until now. Thus, the development of an efficient and cost-effective data search structure that takes into account the tamper-proof nature of blockchain is essential.

To address these requirements, we propose a novel efficient search framework called BPI, which consists of two components: the Blockchain Data Management Forest Plus (BMF+), and the Persistence and Configuration Manager (PCM), which is responsible for managing BMF+ parameters and ensuring data persistence. **The primary contributions are summarized as follows:**

- We propose BMF, an optimized keyword search structure designed for blockchain CR operations, which enables efficient index updates and improves keyword search performance. In addition, it enables arbitrary combinations of composite indexes with virtually no additional overhead. To further enhance efficiency, we introduce Articulated Search, which significantly reduces the search scope and operational overhead.
- We further propose BMF+, which reduces the storage space required by BMF by over 99% in the real Ethereum dataset. Additionally, we introduce PCM to optimize BMF+'s persistence management, ensuring efficient and reliable data maintenance.
- We propose a system model that performs fine-grained validation while minimizing the cost of revalidating the same results. Additionally, we improve the CRC algorithm and use it to achieve a balance between security and communication overhead during the verification process.
- We experimentally demonstrate that BPI exhibits superior search performance and extremely low maintenance overhead in large-scale data set scenarios. Moreover, numerical results also highlight the effectiveness of Articulated Search.

Table 1: Notation definitions.

| Notation | Definition |
| --- | --- |
| $D_i$ | the $i$-th data |
| $R_i$ | the $i$-th query result |
| $Q_a$ | a specific query |
| $VO_{chain}$ | chain-generated validation structures |
| $B_{BMF}$ | max number of children of a node |
| $f_a$ | the mask corresponding to feature $a$ |
| $h_{BMF}$ | height of BMF |
| $len_{BMF}$ | max data capacity of a tree in BMF |
| $k_{BMF}$ | number of BMF features |
| $k_{data_i}$ | number of $data_i$ features |
| $F_{db}(mask)$ | find all offset positions of '1' in the mask |
| $F_{token}(token\ i)$ | calculate where the last query stopped in BMF by token i |
| $DimSize_{data}$ | number of data dimensions |
| $DimSize_i$ | number of features in dimension $i$ |
| $Dim_i$ | the $i$-th dimension |
| $cntLD_i$ | number of nonzero masks in LeafNode at $i$-th dimension |
| $\text{rank}_{\text{row}}(A_i)$ | row rank of $A_i$ |
| $\text{rank}_{\text{col}}(A_i)$ | column rank of $A_i$ |
| $Size_{leaf}$ | size of a leaf node in BMF+ |
| $Size_{middle}$ | size of a middle node in BMF+ |
| $Size_{root}$ | size of a root node in BMF+ |
| $Size_{tree}$ | size of a tree in BMF+ |

## 2 PRELIMINARIES

In this section, we present the preliminaries and articulate the motivation behind our research. Table 1 summarizes the used notations.

### 2.1 Hybrid Storage Blockchain

Blockchain inherently faces storage scalability limitations to maintain network consistency. To address this, Hybrid storage blockchain (HSB) integrates blockchain with off-chain storage [6, 22, 42], storing raw data off-chain while keeping only metadata (e.g., hashes, indices) on-chain. This approach leverages blockchain's immutability and decentralization for security and integrity while benefiting from the high capacity and cost-efficiency of off-chain storage. Additionally, data verification structures like Merkle trees ensure off-chain data integrity, where the Merkle path serves as a Verification Object (VO) to verify the inclusion of queried data.

In Figure 1, both the on-chain smart contract and off-chain Services Provider (SP) maintain an ADS. For each query, the SP generates a $VO_{sp}$ from its ADS to prove data authenticity and integrity. The user verifies this by comparing $VO_{sp}$ with $VO_{chain}$, derived from the ADS maintained by the on-chain smart contract [49].

### 2.2 Merkle Hash Tree

Merkle Hash Tree can be regarded as a type of ADS. As we know, the hash value of a parent node in a Merkle Hash Tree is generated by concatenating and hashing the hash values of its two child nodes. As illustrated in Figure 2, to verify the correctness of $h_6$, we must ascertain its presence in the tree. The constructed VO for this process includes $h_5$, $h_{12}$, $h_{13}$ and $h_{15}$. By hashing together $h_5$ and $h_6$, we obtain $h_{11}$. Subsequently, hashing $h_{11}$ with $h_{12}$ in the VO gives us $h_{14}$. Finally, combining $h_{13}$ with $h_{14}$ yields the root hash. The local computation of the tree's root hash, followed by its comparison with the blockchain's root hash, confirms the presence of $h_6$ if the hashes match, thereby verifying the correctness of $h_6$. However, this verification method requires each data entry to include a substantial amount of redundant proof, leading to significant communication overhead.

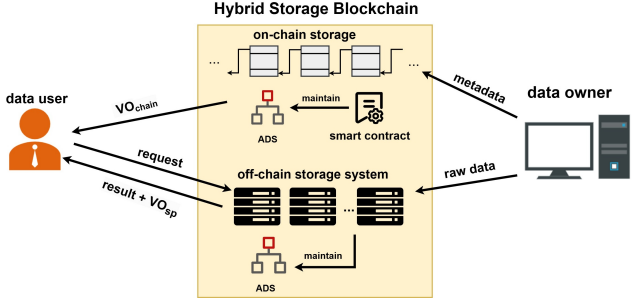

Figure 1: Hybrid storage blockchain.

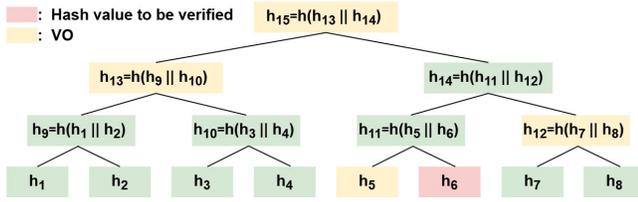

Figure 2: Generation of VO through merkle hash tree.

### 2.3 Verkle Tree

Verkle trees are a cryptographic commitment structure. Unlike Merkle trees, which require logarithmic-sized proofs relative to the number of elements, Verkle trees achieve smaller proof sizes even for larger branching factors, significantly reducing storage and bandwidth overheads. In Verkle trees, verifying whether a value $v_i$ is a child of a node $\text{Node}_v$ reduces to proving that $v_i$ is part of the node's vector of children $(v_1, v_2, \ldots, v_n)$. This is achieved using vector commitments (such as KZG commitments [17]). Specifically, the verifier holds the commitment:

$$C = \text{Commit}(v_1, v_2, \ldots, v_n),$$

while the user generates a proof:

$$\pi_i = \text{Open}(C, i, v_i).$$

The verifier then checks the proof by computing:

$$\text{Verify}(C, i, v_i, \pi_i),$$

which confirms the membership of $v_i$ within the committed vector.

### 2.4 CRC32

CRC32 is a widely used checksum algorithm designed to detect errors during data transmission and storage. It generates a fixed 32-bit checksum, which provides a unique "digital fingerprint" for blocks of data. If the data is modified during transmission or storage, the corresponding CRC value changes, thereby enabling error detection. The core concept of the algorithm involves treating the data as a large polynomial and performing division (using modulo 2 arithmetic) to compute the remainder, which serves as the CRC checksum. In practical applications, the CRC32 algorithm employs a predefined polynomial, known as the generator polynomial, along with an initial value and a final XOR value to compute the CRC checksum for the data.

### 2.5 Problem Formulation

As shown in Figure 1, existing studies often assume the credibility of the $\text{VO}_{chain}$ generated on the blockchain without thoroughly exploring the process by which these credible $\text{VO}_{chain}$ are produced [23, 49]. To ensure reliable verification results, the blockchain relies on multiple nodes participating in the consensus process. These nodes independently obtain verification results prior to reaching a consensus. The simplest method is to allow each node to perform an independent query. However, a key issue is the cost associated with executing these queries, which can significantly impact the overall system efficiency and scalability.

The impact of blockchain's immutability on CRUD operations has been largely overlooked in existing research. In the hybrid storage blockchain model, we assume that the SP stores a data set $DS_0 = \{D_0, D_1, D_2, \ldots, D_k\}$. When the data user issues a query $Q_a$, the result set $RS_0 = \{R_0, R_1, \ldots, R_i\}$ is returned. If we assume that the data set is later added to $\{D_0, D_1, D_2, \ldots, D_k, D_{k+1}, \ldots, D_l\}$. If the data user makes a query $Q_a$ again, the result set extends to $\{R_0, R_1, \ldots, R_i, R_{i+1}, \ldots, R_j\}$. Due to the immutability of blockchain, the original data set $DS_0$ remains unchanged, subsequent queries only need to address the new data $\{D_{k+1}, D_{k+2}, \ldots, D_l\}$ to get $\{R_{i+1}, R_{i+2}, \ldots, R_j\}$, since the previous result set $RS_0$ remains valid.

Existing verification methods primarily rely on Merkle Hash Trees or Verkle Trees [47]. As shown in Figure 2, proof size grows with tree depth and width in Merkle-based verification, while Verkle Trees maintain a fixed proof size. However, both support only coarse-grained verification, requiring the rejection of the entire result upon failure. Fine-grained verification has been explored [23], but users must revalidate results via blockchain for each query.

Inspired by the incremental maintenance approach of LETUS [36], we propose BPI, a lightweight framework for incremental maintenance and search. Based on BPI, we introduce a system model that supports fine-grained validation. Additionally, we improve the CRC algorithm, through which we achieve a balance between communication overhead and verification security. When result verification fails and new query results are obtained, data users can perform local verification without querying the $\text{VO}_{chain}$ on the blockchain.

## 3 SYSTEM MODEL

BPI is a lightweight framework that leverages blockchain immutability. In this configuration, BMF+ collaborates with the existing search engine of Hybrid storage blockchain (HSB), handling and optimizing keyword query operations. Meanwhile, PCM manages the persistence process of BMF+. As illustrated in Figure 3, our system model comprises data owners, an HSB system, and a data user. The HSB system integrates a SP and a blockchain network with smart contract functionality.

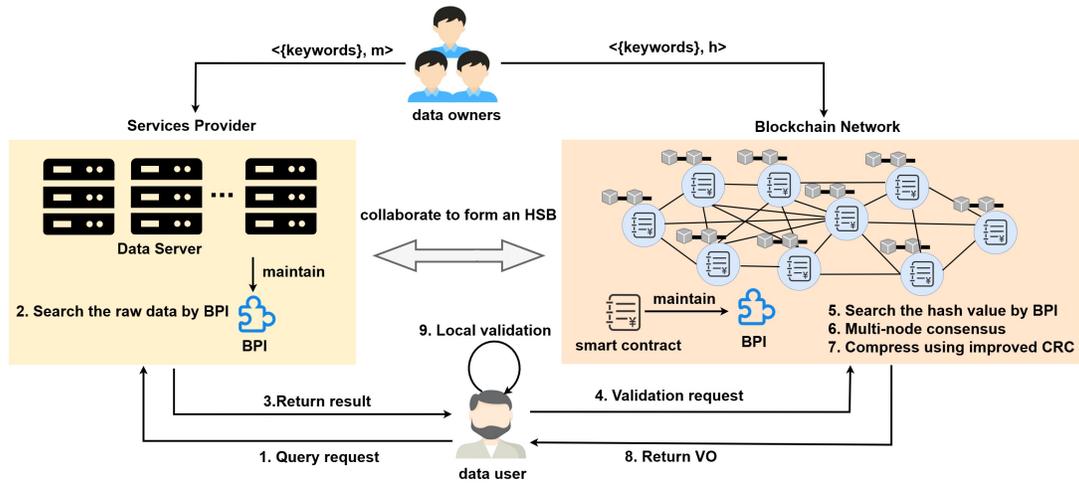

Figure 3: System model.

- **Data owner:** The data owner models the data as a data content $m$, a data digest $h$ (hash value), and a set of keywords $\{keywords\}$. The pair $\langle\{keywords\}, m\rangle$ is outsourced to the SP, while $\langle\{keywords\}, h\rangle$ is uploaded to the blockchain.
- **Services provider:** The SP centralizes substantial computing and storage resources. Original data indices can be obtained through keyword search, facilitating the identification of data that meets the query criteria. Subsequently, the SP packages the data and returns it to the data user.
- **Blockchain network:** The blockchain network is upheld by all network nodes, with a BPI maintained through the blockchain's smart contract. In the validation phase, the user sends a validation request to the blockchain. Miner nodes execute the smart contract, conducting a query to search data digests that fulfill the query conditions. Once consensus is reached among the nodes, the blockchain invokes smart contract to compress the results into a VO, which is subsequently returned to the data user.
- **Data user:** The data user sends a query to the SP. After obtaining the query results, the user can initiate a verification request to the blockchain, which then returns a VO. This VO encapsulates comprehensive information about the digests of accurate query results, allowing the user to verify the completeness and correctness of the query results.

The data owner stores the original data on the Services Provider (SP) and its hash on the blockchain. Both the SP and smart contracts maintain the BPI, updated using $\{keywords\}$, which typically consist of values from all dimensions of the ledger data. The BPI is used to search for original data on the SP, and for hash values on the blockchain. The transaction process is as follows: the data user submits a query to the SP, which processes it and returns the result set $R$. When the user receives the result $R$ returned by the SP, the user locally generates a random value $r$ and sends the query request along with $r$ to the blockchain network. The blockchain executes the query and obtains the hash of the target data through the consensus mechanism. It then passes the hash value and $r$ to the improved CRC, which generates a $k$-bit checksum (we will further elaborate on this in Section 5.5). These checksums form $VO_{\text{chain}}$ and are returned to the user.

The user computes the hash of $R$ and applies the same CRC algorithm to generate $VO_{\text{sp}}$. By comparing $VO_{\text{sp}}$ with $VO_{\text{chain}}$, the user selectively accepts $R$. If the two do not fully match, it indicates that the SP may have been compromised. In this case, the user can query another SP or wait for the current SP to recover. Since the SP does not know the random value $r$, $VO_{\text{chain}}$ remains valid as long as $r$ has not expired. The user can then locally verify a new result $R'$ using the unmatched checksums in $VO_{\text{chain}}$ along with $r$, without sending another verification request to the blockchain.

## 4 BASELINE SOLUTION AND BMF DESIGN

In this section, we initially discuss the search efficiency and maintenance costs associated with B+ trees, which we use as our baseline solution. To tackle the previously mentioned challenges inherent in traditional HSBs, we propose BMF. Our approach considers the distinctive characteristics of blockchain data to effectively maintain data indexes while ensuring high search efficiency.

### 4.1 Baseline Solution

We adopt the B+ tree as the baseline solution for the subsequent discussion. Its query complexity is in $O(\log N)$, where $N$ denotes the number of entries in the index. Maintaining B+ tree indexes during insertions is challenging due to potential structural modifications. Ideally, a key is added to a leaf node without affecting tree height. However, node overflows may necessitate splitting, key redistribution, and tree rebalancing, resulting in an insertion complexity of $O(\log N)$. In blockchain applications with frequent insertions, node splitting and key promotions introduce significant overhead. Moreover, the need to maintain separate indexes for multiple dimensions further increases maintenance costs. The impact of frequent splits can be mitigated through bulk insertions and optimized node allocation strategies.

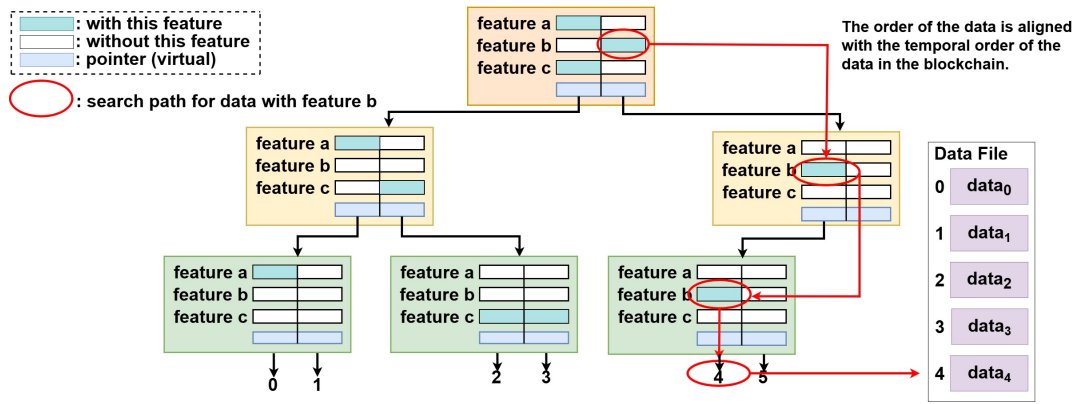

Figure 4: BMF with $B_{BMF}$ of 2 and the search process.

To ensure persistence, indexed search structures must be periodically written to storage. Modern implementations employ page-level updates, buffer management strategies, and write-ahead logging. While these techniques help reduce redundant writes and alleviate maintenance overhead, page-level updates still introduce substantial I/O redundancy, as even minor modifications may require rewriting entire pages.

Table 2: Transaction record (data).

| Dimension | Field Type | Example Value |
|---|---|---|
| from | String | "0x123456789abcdef" |
| to | String | "0x987654321fedcba" |
| toCreate | INT | 1 |
| fromIsContract | INT | 1 |
| toIsContract | INT | 1 |
| value | INT | 10000 |
| gasLimit | INT | 21000 |
| gasPrice | INT | 50000 |
| gasUsed | INT | 15000 |
| callingFunction | String | "transfer" |
| isError | INT | 0 |
| eip2718type | INT | 0 |
| maxFeePerGas | INT | 10000 |
| maxPriorityFeePerGas | INT | 50000 |

## 4.2 BMF Structure Design

In a traditional B+ tree, the order of data in leaf nodes is determined by the key order. In contrast, ledger data on the blockchain is stored in a strict, sequential order that is immutable once created. This characteristic is particularly intriguing as it distinguishes blockchain data from the data structures of traditional B+ trees, which lack immutability. When establishing and maintaining the BMF, we adhere to the principle that the data order in the leaf nodes aligns with the temporal sequence of the blockchain.

Throughout the paper, we use the data format shown in Table 2. A sample keyword query is "find transactions initiated by person_A", which targets records where the keyword "person_A" appears in the first dimension "from". For simplicity, we define such a description, like "transactions initiated by person_A", as a **"feature"** of the data.

Specifically, in this work, BPI monitors all dimensions except the *value* dimension and automatically creates corresponding features for any detected keywords. BPI integrates as a plugin with existing search engines and intercepts keyword queries issued by users for processing. However, if BPI is expected to support range or conditional queries, such as "value > 5,000,000", users need to define custom features.

To optimize space, we use a single bit to indicate whether a data item contains a specific feature. We group $B_{BMF}$ data items (typically 32 or 64, corresponding to `int` or `long long`) into a leaf node, to complete the feature encoding. The resulting encoded sequence is defined as its **mask**. For a feature query, all '1' bits in a leaf node's encoding indicate matching data. Root and middle nodes follow a similar structure: $B_{BMF}$ leaf nodes form a middle node, where each bit denotes whether any child node contains a given feature. This process recurses to the root, where each bit represents the presence of a feature in the corresponding subtree. Unlike B+ trees, which ultimately locate data via pointers, BMF retrieves data through relative offsets, reducing redundant I/O.

Unlike B+ trees, which require frequent modifications to accommodate new data, once a tree in BMF grows into a complete tree, it will no longer change. This is a crucial advantage, as the maintenance cost of a B+ tree index increases with the cumulative data volume. In contrast, within BMF, the maintenance cost impacts only the most recent leaf node that develops. Updates are confined to the latest growing leaf of the most recent tree across the entire forest, identified as the final mask in the feature group.

## 5 WORKFLOW

In this section, we introduce the basic architecture of BMF and then discuss in detail the implementation of BPI in feature creation, update processes, and query workflows, including Articulated Search. We explain how BPI manages data with constant overhead and achieves precise I/O. Finally, we present the core algorithms used during the validation process.

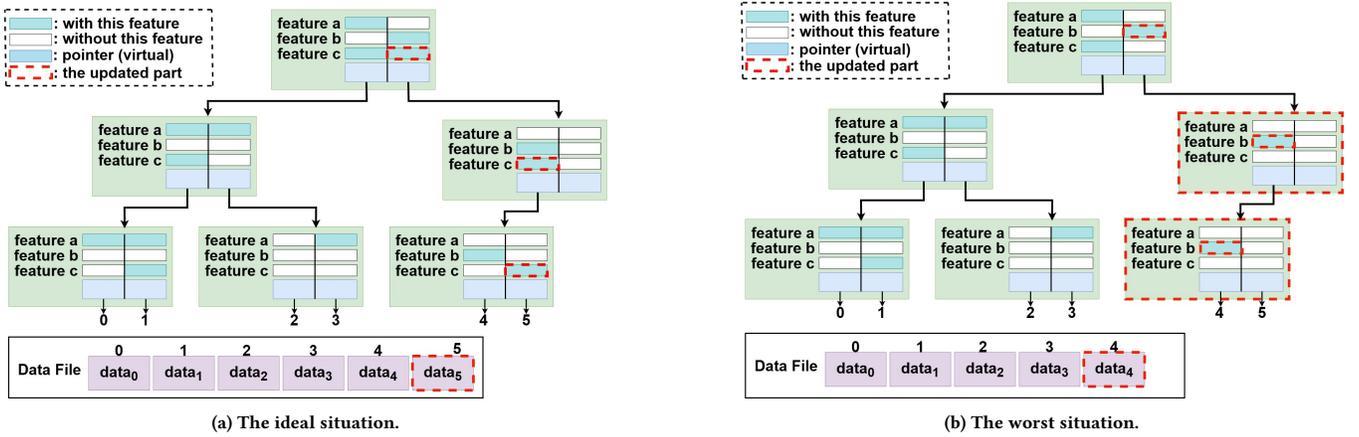

Figure 5: Insert data to a BMF with $B_{BMF}$ of 2.

## 5.1 BMF Parameters

Although the length of a set of masks, denoted by $B_{BMF}$, can be configured as needed, we recommend setting it to 32, which matches the length of an int. This choice allows for direct encoding of the 32 child nodes using an unsignedint, thereby simplifying subsequent data search and update operations. Additionally, a 32-bit mask facilitates the use of the de Bruijn sequence [9] to accelerate bitwise computations, thereby enhancing search efficiency. In the system initialization, $B_{BMF}$ and $h_{BMF}$ are defined, which implies that a tree in BMF can store a maximum of $(B_{BMF})^{h_{BMF}}$ items (i.e. $len_{BMF} = (B_{BMF})^{h_{BMF}}$). In our implementation, we set $B_{BMF} = 32$ and $h_{BMF} = 3$. The three layers of nodes are termed RootNode, MiddleNode, and LeafNode. For ease of illustration in Figure 4 and Figure 5, we set $B_{BMF} = 2$ and the number of keywords $k_{BMF} = 3$ (only within these figures).

## 5.2 Feature Creation in BMF

As shown in Algorithm 1, the creation of features is a bottom-up process that starts with constructing the mask for the LeafNode. For each feature to be created, BMF scans the data and records whether each data entry matches the given feature condition. The result is encoded as a bit sequence where each bit indicates a match or non-match. These bits are grouped into fixed-size blocks to form the leaf buffers. Higher-level buffers (middle and root) are subsequently constructed by aggregating lower-level buffers in the same bitwise manner. Finally, all buffers are stored to files, and an index is generated.

Creating a single feature in BMF incurs relatively high overhead. However, creating multiple features simultaneously introduces little additional overhead compared to a single-feature creation. To improve efficiency, we define a threshold for batch feature creation, $N_{create}$, to ensure that the condition keywords.size() ≥ $N_{create}$ is satisfied when performing the feature creation operation.

---

**Algorithm 1** Create Features in BMF

**Require:** *featureName*, *dimension*, *keywords*
**Ensure:** Create feature bitmaps and store buffers
1: **for each** feature $i$ in *featureName* **do**
2:     Initialize buffers: *rootBuffer*, *middleBuffer*, *leafBuffer*
3:     Initialize bit counter $k \leftarrow 31$, *maskBuffer* $\leftarrow 0$
4:     **for each** data entry $j$ **do**
5:         **if** $data[j][dimension[i]] == keywords[i]$ **then**
6:             *maskBuffer* $\leftarrow$ *maskBuffer* $|$ $(1 \ll k)$
7:         **end if**
8:         $k \leftarrow k - 1$
9:         **if** $k == -1$ **then**
10:            Append *maskBuffer* to *leafBuffer*
11:            Reset $k \leftarrow 31$, *maskBuffer* $\leftarrow 0$
12:         **end if**
13:     **end for**
14:     **if** remaining unprocessed bits exist **then**
15:         Append final *maskBuffer* to *leafBuffer*
16:     **end if**
17:     **for each** buffer $b$ in {*leafBuffer*, *middleBuffer*} **do**
18:         Aggregate $b$ to generate next-level buffer
19:     **end for**
20: **end for**

---

## 5.3 Update in BMF

When new data is added to the blockchain, the baseline solution requires independent updates of the index structures for different dimensions of the data, which may involve adjustments to the tree structure. In contrast, BMF supports for simultaneous updates across all dimensions, and once the tree in BMF is full, it remains immutable. Maintenance operations are limited to the rightmost tree. Figure 4 can be viewed as a forest containing a single tree, while BMF consists of multiple such trees. Despite the presence of multiple trees in BMF, the total number of leaf nodes across all trees remains in the same order of magnitude as the number of leaf nodes in a single tree of the same height in the baseline solution. BMF handles data insertion with constant overhead. In the best case, insertion involves only $k_{data_i} \times h_{BMF}$ iterations of the AND ($\wedge$) operations. In the worst case, the additional overhead is minimal, involving the expansion of only one to three nodes. Figure. 5 (a) and 5 (b) illustrate these scenarios. The constant overhead in data insertion makes BMF highly advantageous for blockchain scenarios where data size continually increases.

When persisting the index data structure to disk, the baseline solution faces challenges in efficiently maintaining file consistency, especially when structural adjustments modify the structure fields of existing index nodes. In contrast, BMF simplifies this process by updating only the latest set of feature masks at each layer. Updates occur under two scenarios: (1) when updating the final bit of each set in the last leaf node's mask while maintaining feature masks from bottom to top (Figure 5 (a)), or (2) when expanding a node, which requires adding a new mask to each set of feature masks across all layers, including additional masks for leaf nodes' feature sets (Figure 5 (b)). The persistence mechanism of BMF+ is described in detail in Section 5.7.

### 5.4 Keyword Search in BMF

Searching with BMF avoids traditional keyword matching and binary search, leveraging bitwise operations to enhance speed. Node search complexity is reduced to $O(1)$ using masking techniques, such as the de Bruijn sequence, to quickly locate the lowest set bit. A recursive search then identifies all data satisfying the specified feature, demonstrating the efficiency of mask arithmetic.

Beyond algorithmic speed, BMF's primary advantage lies in reducing disk I/O, the most time-consuming aspect of searches. In the baseline approach, all data reside in leaf nodes, minimizing I/O but still incurring redundancy. This approach first identifies the page containing the target row, which is then loaded into memory. In contrast, BMF searches a set of offsets at leaf nodes, referencing data positions relative to the first matching entry. By leveraging the immutability and tamper-resistance of blockchain, BMF ensures precise data access while avoiding unnecessary I/O overhead.

In handling queries, which may comprise single or multiple features, a combination like $f_a \wedge f_b$ indicates both $f_a$ and $f_b$ are satisfied. While baseline solutions struggle with cross-dimensional hybrid queries, BMF efficiently manages these with minimal overhead. It achieves this by replacing the mask utilized in the query process with the bitwise result of multiple masks.

Searching follows a top-down process, loading the required mask based on query demands. The search begins by traversing the root node array related to the specified feature. Using $F_{db}(mask)$, the offsets of middle nodes satisfying the condition are computed. This process iterates through middle nodes to leaf nodes, ultimately determining the data's offset in the file. Figure 4 illustrates a BMF tree with $B_{BMF} = 2$ and three features in a database containing five entries. When querying feature $b$, the search starts at the root node, using bitwise operations to navigate to the second node. The process then traverses middle nodes to reach the third leaf node, locating the fourth data entry. In SP, the raw data is searched, while in a blockchain, the data digest serves as the final result.

### 5.5 Validation

Upon completing querying and reaching consensus, the blockchain generates a set of digests for the target results. In our model, we employ the SHA-256 algorithm to create these digests, with each digest occupying 32 bytes. Our objective is to minimize communication overhead through a well-considered design.

We have observed that in verification scenarios with a limited number of results, 32 bits can offer sufficient security. Therefore, we design a dynamic method to adjust the size of the VO according to the number of results. When the number of results is small, the hash values map to a 32-bit space. As the number of results increases, the mapping space expands to 36 bits, 40 bits, 44 bits, and so on, up to a maximum of 128 bits. This mapping is achieved by improving the CRC algorithm. As shown in Algorithm 2, in our design, the generating polynomial is not predefined; instead, it is derived from a 128-bit number that is randomly generated by the data user. This approach effectively reduces the risk of attacks due to polynomial leakage. The specific implementation is as follows: We classify the

---

**Algorithm 2** Improved CRC Calculation($data, k, r$)

**Require:** $data, k, r$
**Ensure:** CRC value
1: Initialize $crc \leftarrow (1 \ll k) - 1$
2: Initialize $polynomial \leftarrow r \wedge crc$
3: **for** each byte $b$ in $data$ **do**
4: $\quad crc \leftarrow crc \oplus b$
5: $\quad$ **for** $i = 0$ to 7 **do**
6: $\quad\quad$ **if** $crc \& 1 \neq 0$ **then**
7: $\quad\quad\quad crc \leftarrow (crc \gg 1) \oplus polynomial$
8: $\quad\quad$ **else**
9: $\quad\quad\quad crc \leftarrow crc \gg 1$
10: $\quad\quad$ **end if**
11: $\quad$ **end for**
12: **end for**
13: **return** $crc \oplus ((1 \ll k) - 1)$

---

potential malicious behaviors of the SP into three categories:

- **Behavior 1 (B1):** The SP mixes malicious data into the result set.
- **Behavior 2 (B2):** The SP conceals certain correct results in the returned result set.
- **Behavior 3 (B3):** The SP returns a result set that is entirely malicious.

Three safety parameters are established during the system initialization: $\alpha$, $\beta$, and $\gamma$. When a data user receives the result set provided by the SP, he may choose to initiate a verification request to the blockchain. The request includes a 128-bit random number $r = r'|c$, here, $c = \sum_{k=8}^{32} 2^{(4k-1)}$, $r'$ is randomly generated by the data user. This ensures that the leading coefficient of the polynomial generated by $r$ is non-zero, thereby reducing the likelihood of collisions to a certain extent. Upon receiving the verification request, the blockchain generates $N_h$ digests that meet the specified conditions and selects an appropriate mapping function to map the hash values to $k$ bits, where $k \in \{a_n \mid a_n = 4n + 28, n \in \mathbb{Z}^+, 1 \leq n \leq 25\}$. It extracts $k$ consecutive bits from $r$ as a polynomial. The criterion for

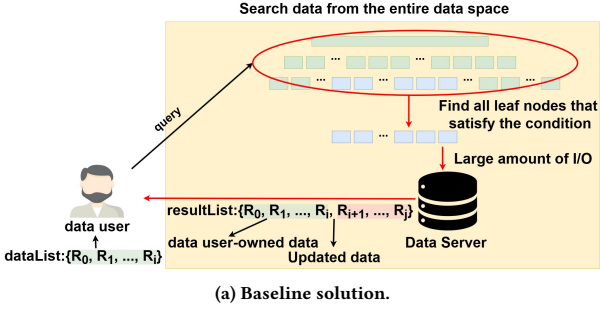
(a) Baseline solution.

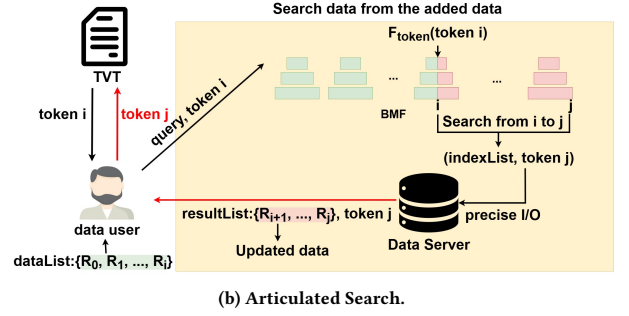
(b) Articulated Search.

Figure 6: Comparison of Articulated Search and baseline solution.

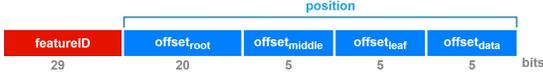

Figure 7: Structure of the token.

choosing $k$ is to ensure that the probability of detecting malicious data under condition **B1** is not less than $\alpha$:
$$1 - \frac{N_h}{2^k} > \alpha.$$

Similarly, the probability of generating a correct error-correcting code under condition **B2** is not less than $\beta$:
$$\prod_{i=2}^{N_h} \left(1 - \frac{i-1}{2^k}\right) > \beta.$$

If $N_h$ is too large to be accommodated by 128 bits, mapping is bypassed, and the hash value is returned directly. The blockchain compresses the result set and returns it as a Verification Object (VO) to the data user. The data user generates a digest from the result set $R$ (of size $N_R$) provided by the SP and selects $k$ for mapping using the same principle. By comparing with the VO, the data user identifies $N_{acc}$ successfully verified data items and $N_h - N_{acc}$ unmatched correction codes, allowing fine-grained verification. If $\frac{N_{acc}}{N_R} < \gamma$, the query is classified under condition **B3**, leading to the rejection of all results to prevent deception by potentially malicious SP data. If $N_h - N_{acc}$ is nonzero, local verification can be performed using new query results, $r$, and correction codes without additional blockchain requests.

### 5.6 Articulated Search

Consider a scenario with 2 billion transactions on Ethereum. A user aims to count those where amounts exceed five million, yielding 100 million matching records. A week later, Ethereum adds 10 million new transactions. The user needs to search only within these new transactions for amounts exceeding the threshold. Reprocessing the initial 2 billion transactions is inefficient and unnecessary due to the immutability of blockchain data, as the user already possesses the relevant records.

Traditional search structures for blockchain scenarios cannot efficiently support this incremental update approach. However, BMF addresses this issue by providing the user with a stop location **token** after each query. When performing the same query again, the user can use this token to resume the search from where they previously stopped, ensuring they only process new data. This approach significantly improves efficiency. The token has a size of 64 bits, as shown in Figure 7, and encapsulates all the necessary information for this query.

As shown in Figure 6a, after a query is issued and the dataset dataList is retrieved, the user may repeat the same query. The baseline solution searches the entire data space for matching index paths each time, returning both newly added data and previously retrieved dataList. However, due to the immutability of blockchain ledger data, this redundant search is unnecessary. In contrast, Articulated Search optimizes this process by issuing a token after each query. The user maintains a **Token Version Table** (TVT) that stores all received tokens. TVT is an ordered list sorted by featureID (Figure 7). By locating the featureID mapped to the current feature in TVT (taking only the most recent one), the system retrieves the latest token returned for the user's previous query on the same feature. When the same query is repeated, the user attaches this token associated with the query.

As shown in Figure 6b, after the initial query, the data user receives dataList and token $i$, which is stored in the TVT. When reissuing the query, the user provides token $i$, enabling BMF to locate the previous query's termination point and restrict the search to the newly added data. This process identifies new matches and returns token $j$, thereby reducing both search and I/O overhead. Even in cases where portions of the locally cached dataList are lost, the missing data can be recovered using prior tokens in the TVT. Thus, the TVT serves as a version management system, allowing for query version rollback.

### 5.7 BMF+

As the number of feature increases, BMF's storage overhead grows proportionally, leading to scalability challenges. To address this, we propose BMF+ and introduce PCM to efficiently manage its persistence and related parameters. We observe that masks with a value of 0 in BMF are never accessed. As shown in Figure 4, when searching for feature $b$, skipping the left subtree has no impact. Thus, BMF+ excludes all masks with a value of 0. PCM maintains a **FirstNode Index Table** (FIT) to record the file offsets of the first nonzero mask in each BMF+ tree's MiddleNode and LeafNode.

Additionally, PCM employs a lightweight **Empty-tree Binary Filter** (EBF), a set of binary filters where each filter corresponds to a specific feature, and each bit within the filter indicates whether the associated BMF+ tree has the specific feature. For 7,000,000 data entries, EBF requires only 214 bits. If a tree does not have this feature, FIT does not store its index pair, as illustrated in Figure 9, where FIT for feature $a$ omits tree#1.

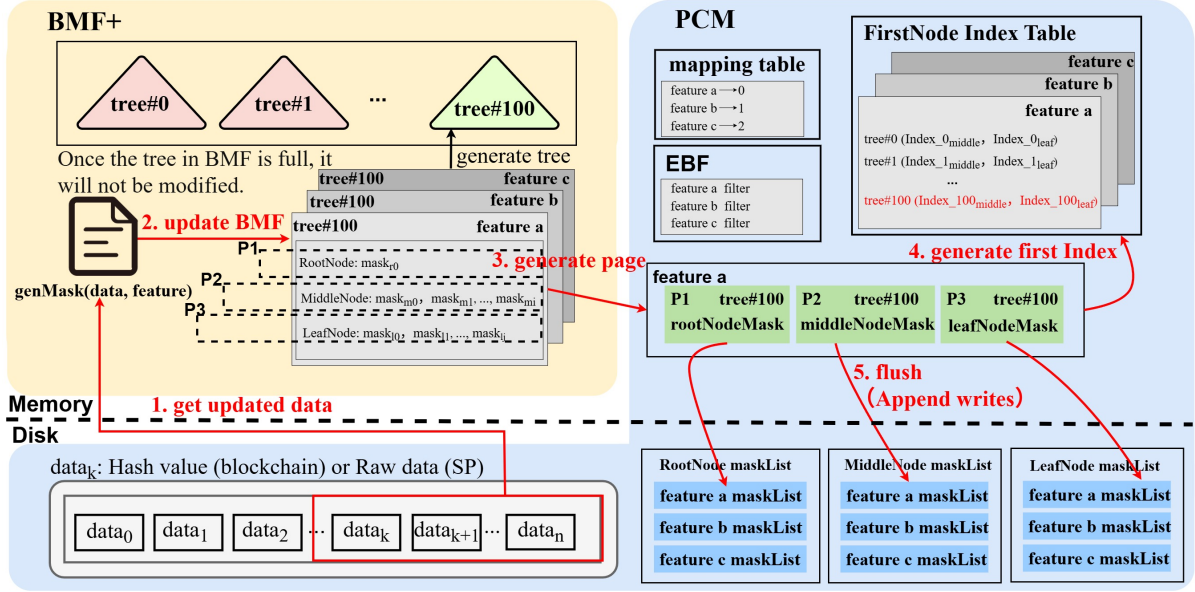

Figure 8: Automatic persistence of BMF+.

In the modeling stage, each data entry has $\text{DimSize}_{\text{data}}$ dimensions, with each dimension's value serving as its feature. Considering a specific dimension $\text{Dim}_i$, the matrix $A_i$ composed of masks in a leaf node, with size $\text{DimSize}_i \times B_{\text{BMF}}$, contains at most one nonzero element per column. Therefore, the number of nonzero masks $\text{cntLD}_i$ satisfies:

$$\text{cntLD}_i = \text{rank}_{\text{row}}(A_i) \leq \text{rank}_{\text{col}}(A_i) \leq B_{\text{BMF}}.$$

The size of a leaf node in BMF+ is bounded by:

$$\text{Size}_{\text{leaf}} \leq B_{\text{BMF}} \cdot B_{\text{BMF}} \cdot \text{DimSize}_{\text{data}} = B_{\text{BMF}}^2 \cdot \text{DimSize}_{\text{data}}.$$

For a middle node's mask, the matrix $A_j$ has at most $\text{cntLD}_i$ nonzero elements per column, leading to:

$$\text{Size}_{\text{middle}} \leq B_{\text{BMF}} \cdot B_{\text{BMF}}^2 \cdot \text{DimSize}_{\text{data}} = B_{\text{BMF}}^3 \cdot \text{DimSize}_{\text{data}}.$$

By extension, we obtain:

$$\text{Size}_{\text{root}} \leq B_{\text{BMF}}^4 \cdot \text{DimSize}_{\text{data}}.$$

Thus, the total size of the index structure for $\text{len}_{\text{BMF}}$ data entries satisfies:

$$\text{Size}_{\text{tree}} \leq \text{Size}_{\text{root}} + B_{\text{BMF}} \cdot \text{Size}_{\text{middle}} + B_{\text{BMF}}^2 \cdot \text{Size}_{\text{leaf}}$$

which simplifies to:

$$\text{Size}_{\text{tree}} \leq 3B_{\text{BMF}}^4 \cdot \text{DimSize}_{\text{data}}.$$

Therefore, regardless of the number of features, the index size per data entry is bounded by:

$$3B_{\text{BMF}} \cdot \text{DimSize}_{\text{data}} \text{ (bits)}.$$

This means that when the number of features per dimension reaches $B_{\text{BMF}}^3$, the size of BMF+ remains constant regardless of feature growth. Experiments in Section 6.4 validate this theoretical bound.

The persistence process includes **manual persistence** and **automatic persistence**. Once BMF+ reaches the end of a tree, automatic persistence is triggered. Figure 8 illustrates PCM's persistence management. First, masks are extracted from the updated data and used to update BMF+. When updates reach the end of tree#100, automatic persistence is initiated. The incremental changes in tree#100 are divided into three pages ($P_1$, $P_2$, $P_3$) and submitted to PCM. PCM updates the FIT and flushes the pages to disk using **append write**. When a new feature is created, a mapping from the feature to its identifier is inserted into the **mapping table**, where the identifier is denoted as $f$ in Algorithm 3.

During manual persistence, the flushed mask on disk may be incomplete. In such cases, subsequent append writes reposition the disk head to the start of the last mask and update it accordingly.

**Example 1.** Assume that the last flushed mask is $mask_{old}$, which was written up to the $m$-th bit, and the first mask in the new page is denoted as $mask_{new}$. Instead of directly writing $mask_{new}$, the system updates the disk content by applying a bitwise OR operation, i.e., $mask_{old} \vee mask_{new}$, and overwriting the previous $mask_{old}$. However, in most cases, persistence simply involves appending the new page content to the end of the file.

### 5.8 Keyword Search in BMF+

As shown in Figure 9, BMF+ removes all zero-value masks, preventing the direct offset-based search as used in BMF. To address this, we propose an enhanced search strategy for BMF+ in this section. As described in Algorithm 3, the query procedure in BMF+ follows a top-down traversal of bitmask trees. Starting at the root, masks corresponding to each feature in the query are loaded and candidate indices are identified. The process continues recursively through the middle and leaf levels, narrowing down the matching entries. Using hierarchical masking and precomputed indexes, BMF+ ensures high-speed keyword matching with minimal I/O.

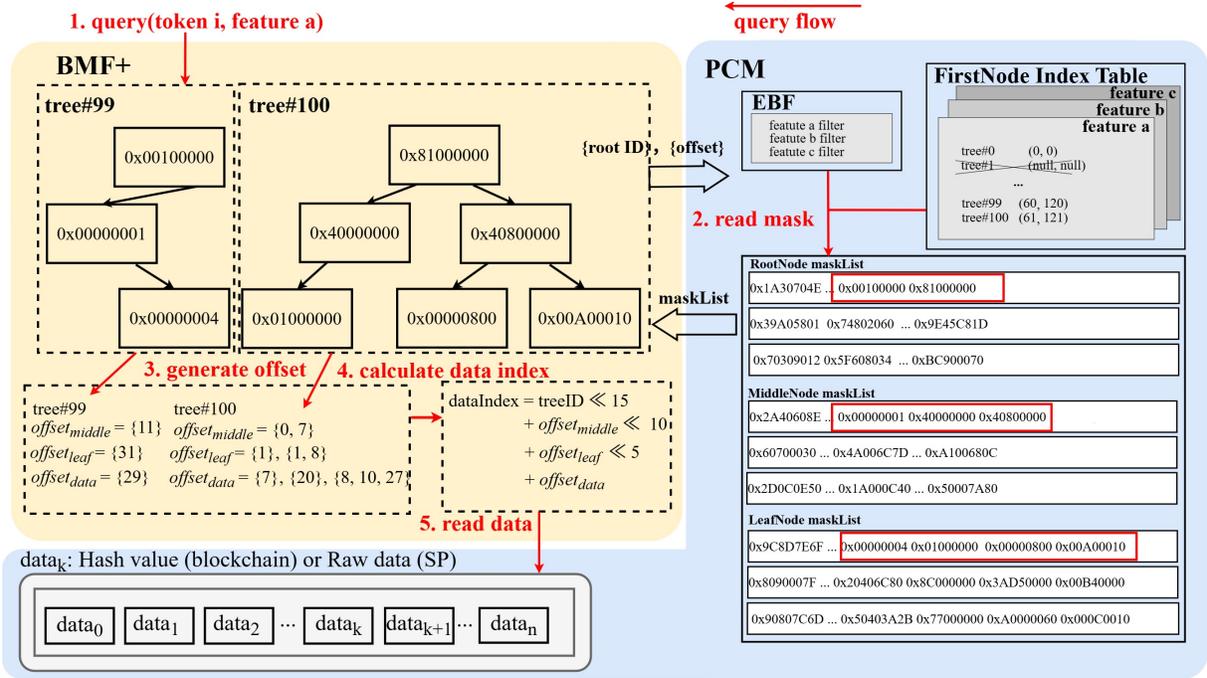

Figure 9: Search in BMF+.

**Example 2.** Figure 9 shows a concrete example. The query process $query(token\ i, feature\ a)$ proceeds as follows.

First, the filter for feature $a$ is searched from the EBF, assuming that 30 occurrences of '1' are identified before reaching the 99-th bit. The 31-st index pair is then read from the FIT, yielding (60, 120). Next, starting from the 31-st mask of feature $a$ in the RootNode maskList, the masks of the root nodes of tree#99 and tree#100 are searched, and the result $\{0x00100000, 0x81000000\}$ is assigned to the temporary variable maskBuffer$_{root}$. Then, execute $F_{db}$(maskBuffer_root), resulting in the $offset_{middle} = \{11\}$ for tree #99 and $offset_{middle} = \{0, 7\}$ for tree #100. It follows that, for feature $a$, tree#99 has empty subtrees except for the 11-th subtree, while tree#100 has empty subtrees except for the 0-th and 7-th subtrees. Next, in the MiddleNode maskList, starting from the 60-th mask of feature $a$, three masks are searched. The first corresponds to the middle node of tree#99, and the next two correspond to the middle nodes of tree#100 (since $\{11\}$ has a size of 1 and $\{0, 7\}$ has a size of 2). This iterative process continues, computing the offsets for the target LeafNode masks, $offset_{leaf}$, and the target data's offset within the LeafNode, $offset_{data}$. The final data offset in the file, denoted as dataIndex, is determined by treeID $\ll \log_2 len_{BMF}$ + $offset_{middle} \ll \log_2 M_n + offset_{leaf} \ll \log_2 L_n + offset_{data}$, where $\ll$ represents the left shift operation. $len_{BMF}$, $M_n$, and $L_n$ denote the data capacities of a tree in BMF+, the data capacity of the middle nodes, and the data capacity of the leaf nodes, respectively.

### 5.9 Feature Creation in BMF+

Typically, each dimension's value in a data record serves as its feature, making feature creation overhead negligible. This is because all preceding masks are set to 0 for any newly introduced feature. Due to BMF+'s incremental maintenance property, it does not require consideration of all previous data. However, to extend BPI's capabilities beyond keyword queries, BPI supports customized conditional queries. For example, transactions with an amount ≥ 5,000,000 can be identified as a feature. In this case, feature creation incurs the complexity described in Section 5.2. Additionally, the size of BMF+ is expanded to $Size_{general} + Size_{define}$, where $Size_{general}$ refers to the size analysis of BMF+ in Section 5.7, and $Size_{define}$ represents the overhead introduced by the customized features.

**Example 3.** Suppose there is a transaction initiated by person_A, who has never appeared as a initiator before. In this case, feature creation is required. In BMF+, since the masks for this new feature are known to be all zeros beforehand, BMF+ skips all prior data and starts building the index from the current record onward.

Customized conditionals mainly support range queries with specific conditions, such as "value > 5,000,000." For a query like "from person_A to person_B," customized conditionals are not needed, such cases only require lightweight bitwise operations, as discussed in Section 5.4 for expressions like $f_a \wedge f_b$.

## 6 EXPERIMENTAL EVALUATION

In this section, we experimentally evaluate the search and insertion efficiency of BPI along with its overall performance. We compare BPI with EthMB+ [10], EASL and two widely used databases: MySQL and PostgreSQL [14]—which utilize B+ tree-based search

**Algorithm 3** Query Algorithm in BMF+

**Require:** *featureList*, a list of feature indices
**Ensure:** Query results in *resultData*
1: **for** $i \leftarrow 0$ **to** theNumberOfRoot $-1$ **do**
2:    Initialize *result* $\leftarrow \emptyset$, *maskBuffer* $\leftarrow \emptyset$
3:    **for each** feature $f$ **in** *featureList* **do**
4:       *maskBuffer*.push(BMFRootNode[$f$][$i$])
5:    **end for**
6:    *tmp0* $\leftarrow$ findAllBitOn(*maskBuffer*)
7:    **for each** $j$ **in** *tmp0* **do**
8:       *maskBuffer* $\leftarrow \emptyset$, *middleIndex* $\leftarrow \emptyset$
9:       **for each** feature $f$ **in** *featureList* **do**
10:          *tmp* $\leftarrow$ getMaskIndex($i, j, -1, f, -1$)
11:          *middleIndex*.push(*tmp*)
12:          *maskBuffer*.push(BMFMiddleNode[$f$][*tmp*])
13:       **end for**
14:       *tmp1* $\leftarrow$ findAllBitOn(*maskBuffer*)
15:       **for each** $k$ **in** *tmp1* **do**
16:          *maskBuffer* $\leftarrow \emptyset$
17:          **for each** feature $f$ **in** *featureList* **do**
18:             *tmp* $\leftarrow$ getMaskIndex(*middleIndex*[$f$])
19:             *maskBuffer*.push(BMFLeafNode[$f$][*tmp*])
20:          **end for**
21:          *tmp2* $\leftarrow$ findAllBitOn(*maskBuffer*)
22:          **for each** $l$ **in** *tmp2* **do**
23:             *result*.push$\big((i \ll 15) + (j \ll 10) + (k \ll 5) + l\big)$
24:          **end for**
25:       **end for**
26:    **end for**
27: **end for**
28: readDataForQuery(*dataFileName*, *result*, *resultData*)

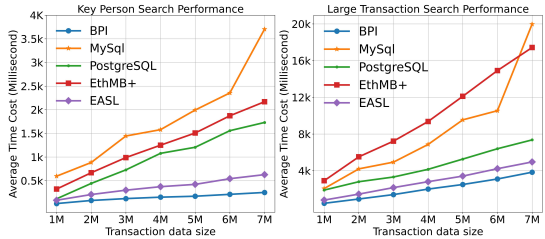

**Figure 10: Basic search performance.**

structures. These databases are commonly employed for Services Provider (SP) and blockchain search tasks due to their efficient query capabilities [29, 34]. We conduct five sets of comparative experiments: (1) basic search efficiency, (2) maintenance overhead for large-scale data insertions, (3) effectiveness of Articulated Search, (4) BMF+ optimization effect, and (5) impact of the improved CRC scheme. We collect transaction data from the Ethereum network, including 7, 305, 457 records from block 2,000,000 to 2,140,994 [51]. We define two features: the transaction initiator "0x32be343b94f860124d c4fee278fdcbd38c102d88" (feature name: Key Person) and transactions with amounts $\geq 5,000,000$ (feature name: Large Transactions). Our test environment is a computer with an 8-core i5-12450H CPU, 32 GB RAM, and Windows 11. All algorithms are implemented in C++. Database Management Systems (DBMSs) build indexes on the "id", "from", and "value" dimensions, with "id" serving as a clustered primary key to simulate temporal ordering. Composite indexes are further created on {id, from} and {id, value}. EthMB+ constructs the same set of indexes but stores hash values only in the index nodes on the "id" field. EASL is primarily implemented using skip lists and fully leverages blockchain immutability to track relationships between data across different blocks, enabling version tracing. In our experiments, we use "id" as the primary key, with each node storing both "value" and "from" dimension values, thereby approximating the functionality of composite indexes. In addition, we build direct indexes on the "from" and "value" dimensions.

### 6.1 Analysis for Basic Data Search

We conduct experiments across seven groups, with data sizes ranging from one million (1M) to seven million (7M) records, in equal increments, and average the results over 100 experiments for each group. The cache is cleared before each experiment to eliminate any caching effects on the results.

Figure 10 (left) presents the results of the Key Person search, where 67,760 transactions are identified from over 7 million records. When feature data points occupy a relatively small proportion of the entire query space, BPI significantly outperforms the four baseline methods on large datasets, requiring far fewer I/O operations. As the dataset grows, the dispersion of feature increases redundant I/O in the baselines. Figure 10 (right) shows the Large Transactions query, identifying 1,220,659 entries, about one-sixth of the dataset. When feature data is densely distributed, baseline methods experience fewer redundant I/O operations since relevant pages are often preloaded into memory. Even in such cases, BPI maintains a notable advantage. In the 7M dataset, BPI completes the Key Person search in 250 *ms*, while PostgreSQL, EthMB+ and EASL take 1726 *ms*, 2169 *ms*, and 627 *ms*, respectively, achieving at least **2.5×** improvement. For the Large Transactions query, BPI records 3838 *ms*, whereas PostgreSQL, EthMB+ and EASL require 7552 *ms*, 16291 *ms* and 4656 *ms*, respectively, resulting in at least **1.2×** speedup. Compared to BPI, the baselines incur additional overhead mainly from redundant I/O, while EASL also suffers from in-memory computation and table lookup costs.

Overall, BPI's search performance exceeds that of the baseline solutions, especially when the search results are dispersed across the dataset. Its search efficiency can be **2.5×** to **20×** greater, or even more, primarily due to its ability to leverage the immutable nature of blockchain data to avoid redundant I/O operations. Even when the result data is more centralized within the dataset, BPI's search speed remains approximately **1.2×** faster than that of the baseline solutions.

### 6.2 Analysis for Data Insertion

Each experiment uses a database with 1 to 7 million entries, inserting 10,000 new records per batch. The average insertion time, including index persistence, is recorded over multiple runs. Like search operations, insertion overhead is dominated by I/O. EASL exploits the "append-only" property when indexing temporally increasing keys (e.g., "id"), but faces similar challenges to DBMSs for non-temporal keys such as "from" or "value". In contrast, BPI enables fixed and minimal overhead for both persistence and indexing. As shown in Figure 5(a), BPI incurs minimal I/O cost and maintains a constant insertion time of ~40 $ms$ per 10,000 entries, a key advantage in blockchain scenarios with ever-growing data. Table 3 shows that BPI is at least **30×** faster.

Table 3: Insertion performance.

| Data Size | Database Model | | | | |
|---|---|---|---|---|---|
| | BPI | EthMB+ | MySQL | EASL | PostgreSQL |
| 1M | 54 $ms$ | 3129 $ms$ | 5820 $ms$ | 1521 $ms$ | 3780 $ms$ |
| 2M | 21 $ms$ | 4110 $ms$ | 11027 $ms$ | 2962 $ms$ | 5605 $ms$ |
| 3M | 21 $ms$ | 6587 $ms$ | 14446 $ms$ | 3455 $ms$ | 6632 $ms$ |
| 4M | 24 $ms$ | 9021 $ms$ | 18730 $ms$ | 4201 $ms$ | 8308 $ms$ |
| 5M | 26 $ms$ | 10475 $ms$ | 24218 $ms$ | 4824 $ms$ | 10538 $ms$ |
| 6M | 27 $ms$ | 13002 $ms$ | 29828 $ms$ | 6269 $ms$ | 12793 $ms$ |
| 7M | 28 $ms$ | 15393 $ms$ | 34848 $ms$ | 7463 $ms$ | 14344 $ms$ |

Table 4: Performance of Articulated Search on key person.

| Data Size | Database Model | | | | |
|---|---|---|---|---|---|
| | BPI | EthMB+ | MySQL | PostgreSQL | EASL |
| 1M | 1.68 $ms$ | 32.84 $ms$ | 40.85 $ms$ | 10.85 $ms$ | 7.55 $ms$ |
| 2M | 1.72 $ms$ | 36.21 $ms$ | 41.10 $ms$ | 12.03 $ms$ | 7.78 $ms$ |
| 3M | 1.80 $ms$ | 42.12 $ms$ | 40.95 $ms$ | 11.75 $ms$ | 7.05 $ms$ |
| 4M | 1.67 $ms$ | 42.50 $ms$ | 44.23 $ms$ | 12.85 $ms$ | 7.15 $ms$ |
| 5M | 1.74 $ms$ | 38.08 $ms$ | 42.15 $ms$ | 11.48 $ms$ | 7.50 $ms$ |
| 6M | 1.78 $ms$ | 42.37 $ms$ | 42.02 $ms$ | 11.97 $ms$ | 7.98 $ms$ |
| 7M | 1.67 $ms$ | 41.81 $ms$ | 43.52 $ms$ | 11.95 $ms$ | 7.55 $ms$ |

Table 5: Performance of Articulated Search on large transactions.

| Data Size | Database Model | | | | |
|---|---|---|---|---|---|
| | BPI | EthMB+ | MySQL | PostgreSQL | EASL |
| 1M | 24 $ms$ | 183.17 $ms$ | 196.45 $ms$ | 52.61 $ms$ | 162.51 $ms$ |
| 2M | 26 $ms$ | 202.41 $ms$ | 210.15 $ms$ | 58.32 $ms$ | 157.05 $ms$ |
| 3M | 27 $ms$ | 194.68 $ms$ | 221.95 $ms$ | 62.73 $ms$ | 162.51 $ms$ |
| 4M | 25 $ms$ | 197.29 $ms$ | 246.27 $ms$ | 62.25 $ms$ | 152.73 $ms$ |
| 5M | 23 $ms$ | 210.48 $ms$ | 250.56 $ms$ | 51.82 $ms$ | 144.65 $ms$ |
| 6M | 27 $ms$ | 208.49 $ms$ | 240.94 $ms$ | 71.15 $ms$ | 148.82 $ms$ |
| 7M | 28 $ms$ | 199.62 $ms$ | 241.31 $ms$ | 71.82 $ms$ | 147.99 $ms$ |

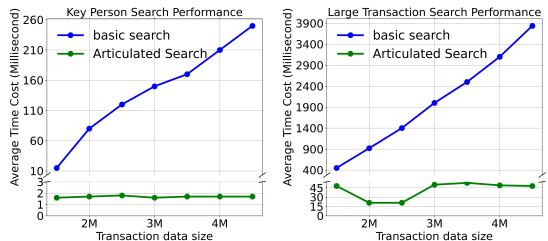

Figure 11: Comparison between basic search and Articulated Search.

### 6.3 Analysis for Articulated Search

In this section, we conduct seven experiments with datasets ranging from 1M to 7M. Each inserts 10,000 new entries, followed by searches for recent transactions initiated by the Key Person or classified as Large Transactions. Results are averaged over 100 iterations for accuracy. MySQL, PostgreSQL and EthMB+ detect relevant transactions using composite indexes, checking timestamps beyond the last recorded entry before insertion. EASL first locates the target timestamps, then verifies the values of the "from" and "value" fields, and finally retrieves the corresponding data. In contrast, BPI employs Articulated Search to efficiently retrieve the most recent 10,000 matching entries. As shown in Tables 4 and 5, baseline methods reduce query latency by paying an upfront cost to build composite indexes. However, additional table lookups still result in query overheads that are **2 to 20×** higher than BPI. EASL performs competitively, but it still incurs additional costs to filter records beyond the target timestamp, leading to a query overhead approximately **4×** that of BPI. These results highlight the efficiency of the BPI.

Figure 11 compares the time overhead of BPI with and without Articulated Search (basic search) under two types of queries. Clearly, Articulated Search is highly effective for large-scale blockchain data, as its overhead remains unchanged as data size grows.

Table 6: Feature statistics of experimental data.

| Dimension | Feature Count |
|---|---|
| *from* | 348726 |
| *to* | 336980 |
| *toCreate* | 56474 |
| *fromIsContract* | 2 |
| *toIsContract* | 2 |
| *value* | 1689 |
| *gasLimit* | 35179 |
| *gasPrice* | 204560 |
| *gasUsed* | 39505 |
| *callingFunction* | 42871 |
| *isError* | 2 |
| *eip2718type* | 2 |
| *maxFeePerGas* | 2 |
| *maxPriorityFeePerGas* | 2 |

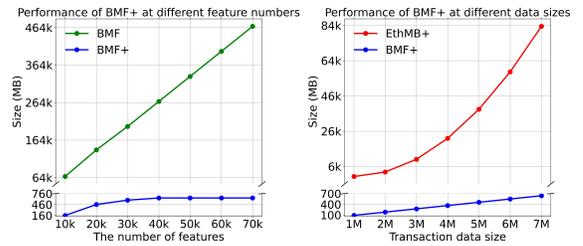

Figure 12: Performance of BMF+.

### 6.4 Analysis for BMF+

In Table 6, the experimental dataset consists of 1,065,996 features, with six Boolean dimensions. EthMB+ builds indexes only for non-Boolean dimensions. The "Value" dimension includes features with at least 100 occurrences, while all other dimensions retain every occurrence.

This section presents two sets of experiments. The first set configures the dataset with 10,000, 20,000, ..., and 70,000 features per non-Boolean dimension to evaluate BMF+'s impact on structural size optimization compared to BMF. The second set examines the structural size of BMF+ and EthMB+ using $1M$, $2M$, ..., and $7M$ data records, covering all 1,065,996 features. Experimental results show that BMF+ effectively curtails memory growth as the number of features increases.

As shown in Figure 12, the size of BMF grows linearly with additional features under a fixed data volume. In contrast, BMF+ incorporates structural optimizations. Once each dimension accumulates $(B_{BMF})^3$ features, resulting in a total of $8(B_{BMF})^3$ features, the size of BMF+ remains stable thereafter. At a dataset scale of 7M records, BMF+ requires at most 640MB. In the full-data experiment, BMF+ reduces storage overhead by over 99% compared to BMF. Furthermore, BMF+ significantly outperforms EthMB+, whose storage usage scales rapidly with data volume due to accumulated hash values. While BMF+ maintains a fixed 640MB footprint at 7M records, EthMB+'s single-dimension index reaches 800MB at just 2.5M records. In addition, the storage overhead of both DBMSs and EASL increases proportionally with the number of indexes created. Under the indexing configurations introduced earlier, on the 7M dataset, DBMS incurs approximately 1500MB of storage overhead, while EASL reaches around 4700MB. In real-world production settings, both typically require additional indexes to support a wide range of query predicates. These results underscore the scalability of BMF+, which minimizes storage overhead without relying on auxiliary indexes. Although BMF+ is less efficient than DBMSs for range queries, it supports conjunctive queries over arbitrary keyword sets without introducing extra index costs.

Table 7: Validation performance.

| Algorithm | Data Size of 10,000 | | |
|---|---|---|---|
| | cost | $\alpha$ | $\beta$ |
| CRC32 | 20.1 ms | $1 - 10^{-5}$ | $e^{-0.01164}$ |
| CRC64 | 84.96 ms | $1 - 10^{-15}$ | $e^{-2.71 \times 10^{-12}}$ |
| CRC128 | 268.01 ms | $1 - 10^{-34}$ | $e^{-1.47 \times 10^{-20}}$ |

### 6.5 Analysis for CRC-based Validation Scheme

In our study, we set up a local Ethereum test network and implement the BPI. During the verification process, a round of consensus and a round of mapping follow the blockchain network query. This requires the data user to perform one round of hashing, one round of mapping, and one round of comparison. In our experimental design, $\alpha$ and $\beta$ are no longer preset values. We test the CRC algorithm with $N_h = 10,000$ (result set size), setting the mapping space to three levels: 32, 64, and 128. We evaluate the verification time required and compute the achievable levels of security parameters $\alpha$ and $\beta$. The experimental results indicate that the average time required for one round of hashing is 2.2 ms, one round of comparison takes 1.1 ms, and the time required for one round of mapping is shown in Table 7. The experimental results demonstrate that our system design reduces communication overhead by 50% to 87.5% during the verification process, while maintaining satisfactory levels of security and efficiency.

## 7 LIMITATION DISCUSSION

BPI significantly enhances search performance and maintenance efficiency but is restricted to keyword searches and may not fully cover all search scenarios. No single search structure perfectly supports all request types, and BPI is no exception [19, 24]. Our objective is to complement rather than replace existing search structures. BPI is particularly effective in scenarios requiring high search performance and low maintenance costs while uniquely supporting Articulated Search.

Specifically, BPI faces challenges in conditional and range searches due to its lack of support for arbitrary interval queries, as it relies on predefined independent keywords instead. For example, if $f_1$ defines transaction records under 1000 and $f_2$ defines records greater than 5000, transactions between 1000 and 5000 can be represented as $(\neg f_1) \wedge (\neg f_2)$. Overall, BPI aligns well with the tamper-proof, chronological nature of blockchain data. It offers minimal search performance and maintenance overhead, making it a lightweight search framework. Moreover, BPI supports Articulated Search at an ultra-low cost and demonstrates excellent storage efficiency.

## 8 RELATED WORK

Blockchain technology, known for its decentralization, transparency, security, traceability, and smart contract support, provides a solid foundation for distributed data storage and transaction verification [4, 5, 25, 52, 53]. However, traditional blockchain systems face scalability and storage constraints, limiting broader adoption [16, 33, 40]. HSBs address these issues by integrating on-chain and off-chain storage, improving processing speed, increasing capacity, and reducing transaction costs [1, 8, 23, 32, 42, 49, 54]. These solutions combine blockchain's security with the scalability of off-chain storage [12, 27].

Despite these advances, the rapid growth and immutability of blockchain data pose ongoing challenges. Traditional databases, designed for CRUD operations, are not well suited to blockchain's append-only nature, resulting in inefficient queries and insertions as data accumulates [16, 33, 40]. While prior work has improved blockchain search performance [43, 46, 48], it often overlooks the implications of immutability, which leads to increased search and maintenance overhead.

Structures tailored for append-only databases, such as EASL [30] and log-structured merge-tree (LSM-tree) [31, 38], achieve strong point and range query performance but face limitations for queries on non-time-ordered keywords. Bitmap indexes [37] and Bloom filters (such as Bloofi[7]) offer high performance and can better support composite queries, yet they require additional bits to represent each query condition, resulting in considerable storage overhead. Many techniques address this by leveraging bitmap sparsity [2, 41]. However, sparsity-based compression, though effective in reducing space, introduces difficulty in decompression for arbitrary ranges. For example, when querying transactions within a specific time interval, such schemes struggle to quickly and precisely identify the corresponding data segments.

CRC is a widely used error-detection method that computes checksums via polynomial division. While hash remapping reduces communication overhead, traditional CRC outputs are fixed-length [13, 18, 28]. Our model adapts CRC to produce variable-length codes based on result set size, reducing cost while preserving security.

Moreover, ADS-based query verification remains vulnerable to malicious SPs, compromising result integrity [1, 8]. BPI fully leverages blockchain immutability to minimize I/O overhead and reduce space usage in large-scale datasets. Its system model guarantees query completeness and correctness while significantly lowering redundant verification costs. In our scenario, many masks are `0x00000000`. We achieve compression by skipping operations on these zero masks and avoiding their storage, which significantly reduces storage overhead while still supporting efficient random access.

## 9 CONCLUSION

In this paper, we address the unique data search challenges in Hybrid storage blockchains (HSBs) systems, stemming from blockchain's immutability. We propose **BPI**, a novel lightweight framework that leverages the temporal order inherent in blockchain data to enable efficient keyword search while minimizing overhead during update operations. Additionally, we develop Articulated Search to narrow the search space, improving efficiency while reducing computational load. Our method ensures comprehensive verification via blockchain without redundant validation and introduces an enhanced CRC algorithm for VO compression, significantly lowering communication overhead. Experimental results confirm BPI's superior search performance with lower maintenance costs, offering advantages in keyword search over SP's commonly used search engines and Ethereum's EthMB+, making it well-suited for large-scale blockchain applications.

## ACKNOWLEDGMENT

This work was supported in part by the National Science Foundation of China under Grants (62472375, 62125206 and 62502446), and in part by the Major Program of National Natural Science Foundation of Zhejiang (LD24F020014, LD25F020002), and in part by the Zhejiang Pioneer (Jianbing) Project (2024C01032), in part by the Ningbo Yongjiang Talent Programme (2023A-198-G and 2024A-402-G), and in part by the fund of Oak Grove Ventures-School of Software Technology, Zhejiang University Blockchain Joint Lab.